\documentclass[amsmath,amssymb,twocolumn,pra]{revtex4-1}
\usepackage{braket,hyperref,graphicx}
\pdfoutput=1

\newcommand{\HS}{\mathcal{H}}
\newcommand{\R}{\mathbb{R}}
\newcommand{\idn}[1]{I_{#1}}
\DeclareMathOperator{\tr}{Tr}

\hypersetup{pdfauthor={Matthew F. Pusey},pdftitle={Negativity and steering: a stronger Peres conjecture}}
\begin{document}
\title{Negativity and steering: a stronger Peres conjecture}
\date{May 8, 2013}

\author{Matthew F. Pusey}
\email{m@physics.org}
\affiliation{Department of Physics, Imperial College London, Prince Consort Road, London SW7 2AZ, United Kingdom}

\begin{abstract}
  The violation of a Bell inequality certifies the presence of entanglement even if neither party trusts their measurement devices. Recently Moroder \textit{et. al.} showed how to make this statement quantitative, using semidefinite programming to calculate how much entanglement is certified by a given violation. Here I adapt their techniques to the case where Bob's measurement devices are in fact trusted, the setting for ``EPR-steering'' inequalities. Interestingly, all of the steering inequalities studied turn out to require negativity for their violation. This supports a significant strengthening of Peres' conjecture that negativity is required to violate a bipartite Bell inequality.
\end{abstract}
\maketitle

Entanglement \cite{entangrev} seems to lie at the heart of both the mysteries and the applications of quantum theory. Its quantification by various entanglement measures is therefore important. Suppose that Alice and Bob receive many copies of some quantum state. If they both have access to suitable trusted measurement devices, they can perform ``local tomography'', reconstructing the density matrix $\rho_{AB}$. This, in turn, can be used to calculate entanglement measures, such as the negativity \cite{neg}, defined as the total magnitude of the negative eigenvalues of $\rho_{AB}^{T_A}$.

However, Alice and Bob may not trust their measuring devices, and therefore cannot rely on the correctness of any reconstructed $\rho_{AB}$. Nevertheless, they can still estimate the probabilities $p(a,b | x,y)$ of getting outcomes $(a,b)$ when they choose the measurements $(x,y)$. If these probabilities violate a Bell inequality (and Alice and Bob believe their measurement devices are unable to communicate), they can be certain that the state is entangled. Moroder \textit{et. al.} \cite{deq} have recently shown how the magnitude of that Bell violation can furthermore be used to calculate a lower bound on the negativity.

Not all entangled states can violate a Bell inequality \cite{werner}. Therefore it may be useful to study the intermediate case where Alice does not have trusted measuring devices, and yet Bob does. This is known as the ``EPR-steering'' scenario \cite{steer}. In this case Bob can do state tomography on his system, and the parties can then estimate $\sigma_{a|x}$, the collapsed or steered state for Bob that is found when Alice gets outcome $a$ from measurement $x$. If the $\sigma_{a|x}$ violate a ``steering inequality'', then their bipartite state is entangled, and I will show, for the simplest class of steering inequalities, how to calculate lower bounds on the negativity for a given violation. The results suggest a strengthening of the long-standing Peres conjecture \cite{peresconjecture}.

\section{EPR Steering: recap and notation}
Suppose Alice can choose between $m_A$ measurement settings, each of which can result in one of $n_A$ outcomes (all of the following can trivially be adapted to the case when different measurement settings have different numbers of outcomes). Suppose Bob has a $d_B$-dimensional quantum system. Define an ``assemblage'' to be a set of $d_B \times d_B$ Hermitian matrices $\sigma_{a|x}$ where $a$ ranges from $1$ to $n_A$ and $x$ ranges from $1$ to $m_A$. We require the $\sigma_{a|x}$ to be positive and $\sum_a \sigma_{a|x}$ to be independent of $x$ and trace 1. We do not require the $\sigma_{a|x}$ to be normalized, instead $\tr(\sigma_{a|x})$ gives the probability that if Alice performs measurement $x$ she obtains outcome $a$, whilst $\sigma_{a|x} / \tr(\sigma_{a|x})$ is the resulting state on Bob's system.

Does the dependence of Bob's state on Alice's measurement results represent ``spooky action at a distance''? Not if there is a set of normalized states $\sigma_\lambda$ with probability distributions $p(\lambda)$ and $p(a|\lambda, x)$ such that $\sigma_{a|x} = \sum_\lambda p(\lambda)p(a|\lambda,x)\sigma_\lambda$. In that case, we can comfort ourselves that Bob's system was in some fixed state $\sigma_\lambda$ all along, and Alice's measurement outcome simply gave us classical information about $\lambda$, causing us to update our probability distribution for it from $p(\lambda)$ to $p(\lambda|a,x) = p(a|\lambda,x)p(\lambda) / p(a|x)$ and therefore assign the state $\sigma_{a|x}/p(a|x)$ to Bob. This is called a local hidden state (LHS) model, and the lack of such a model for some assemblages is called ``steering'' \cite{steer}, taken to be the formal definition of an EPR paradox.

The classic example is Bohm's qubit reformulation \cite{bohm} of the original EPR \cite{epr} setup. This has $n_A = m_A = d_b = 2$, with $\sigma_{1|1} = \ket{0}\bra{0}/2, \sigma_{2|1} = \ket{1}\bra{1}/2, \sigma_{1|2} = \ket{+}\bra{+}/2$ and $\sigma_{2|2} = \ket{-}\bra{-}/2$. This can trivially be seen to lack a LHS model, because pure states cannot be decomposed into any other states.

\begin{table*}
\renewcommand{\arraystretch}{1.5}
\begin{tabular}{|p{3cm}|p{3.6cm}|p{5.1cm}|p{5.5cm}|}
  \hline
 Scenario  & Tomography & Steering & Bell nonlocality \\
  \hline
Trusted parties & Both & Bob & Neither \\
Key parameters & Dimensions $d_A, d_B$ & Settings $m_A$, outcomes $n_A$, dim. $d_B$ & Settings $m_A, m_B$, outcomes $n_A, n_B$\\
Data & $\rho_{AB} \in L(\HS^{d_Ad_B})$, state & $\sigma_{a|x} \in L(\HS^{d_B})$, ``assemblage'' & $p(a,b|x,y) \in \R$, probabilities \\
\hline
Positive & $\rho_{AB} \geq 0$ & $\sigma_{a|x} \geq 0 \quad\forall a,x$ & $p(a,b|x,y) \geq 0 \quad\forall a,b,x,y$ \\
Normalized & $\tr(\rho_{AB}) = 1$ & $\sum_a \tr(\sigma_{a|x}) = 1 \quad\forall x$ & $\sum_{a,b} p(a,b|x,y) = 1 \quad\forall x,y$ \\
No-signalling $A\to B$ & Implicit & $\sum_a \sigma_{a|x}$ independent of $x$ & $\sum_a p(a,b|x,y)$ independent of $x$ \\
No-signalling $B\to A$ & Implicit & Implicit & $\sum_b p(a,b|x,y)$ independent of $y$ \\
Allowed in QM & Whenever above satisfied & Whenever above satisfied \cite{schr,HJW} & It's complicated, see e.g. \cite{hierarchy}\\
Creatable using local operations \& shared randomness& $\rho_{AB} = \sum_\lambda p(\lambda) \rho_\lambda \otimes \sigma_\lambda$ Satisfies all entanglement witnessess, is ``separable'' (Hard to check in general)& $\sigma_{a|x} = \sum_\lambda p(\lambda) p(a|x,\lambda) \sigma_\lambda$\newline Satisfies all steering inequalities, has ``local hidden state (LHS) model'' (Checkable with SDP) & $p(a,b|x,y) = \sum_\lambda p(\lambda)p(a|x,\lambda)p(b|y,\lambda)$ Satisfies all Bell inequalities, has ``local hidden variables (LHV) model'' (Checkable with linear program)\\
\hline
\end{tabular}
\caption{Summary of three scenarios in which bipartite entanglement can be quantified. By choosing POVMs $E_{a|x}$ for Alice one can turn a state $\rho_{AB}$ into an assemblage $\sigma_{a|x} = \tr_A((E_{a|x}\otimes \idn{B}) \rho_{AB})$. By choosing POVMs $E_{b|y}$ for Bob one can turn an assemblage $\sigma_{a|x}$ into a probabilities $p(a,b|x,y) = \tr(E_{b|y}\sigma_{a|x})$. These mappings preserve all the listed properties, in particular a separable state always provides an LHS model, which in turn always provides an LHV model. On the other hand, by encoding Bob's classical data using computational basis states, an LHV model can always be turned into an LHS model with particular measurements for Bob, which can similarly be turned into a separable state with particular measurements for Alice. Combining both directions we see that an assemblage could have arisen from a separable state if and only if it has an LHS model.}
\label{scenariotable}
\end{table*}

But can this assemblage be realised in quantum mechanics? Yes: Bohm gave an explicit two-qubit entangled state $\rho_{AB}$ and measurements $E_{a|x}$ for Alice that achieve it, i.e. $\sigma_{a|x} = \tr_A((E_{a|x}\otimes \idn{B}) \rho_{AB})$. However it is not necessary to check this, because Schr\"odinger \cite{schr} (and later HJW \cite{HJW}, among others) have shown that \emph{any} assemblage satisfying the basic criteria given above has a quantum realisation. However, that result makes use of a pure entangled state between Alice and Bob. The aim of this paper is explore to what extent we can get by with less entanglement than that.

\section{Steering Inequalities: A semidefinite warm-up}

Let $X$ be a hermitian matrix. A semidefinite program \cite{sdp} is the minimization of some linear functional of $X$ subject to $X \geq 0$ and bounds on linear functionals of $X$. We can easily generalise this to multiple $X_i$ by constructing a block-diagonal $X$ containing each one. Semi-definite programs can be solved in polynomial time using freely available code, e.g. \cite{sdpt,csdp}.

For a given $n_A, m_A, d_B$, define a ``steering functional'' $F$ by a set of $d_B \times d_B$ Hermitian matrices $F_{a|x}$ where $a$ ranges from $1$ to $n_A$ and $x$ ranges from $1$ to $m_A$. $F$ maps an assemblage  to a real number by $\sum_{a,x} \tr(F_{a|x}\sigma_{a|x})$. (Recall that any linear map from the Hermitian matrices to the real numbers can be written $\tr(F\cdot)$ for some $F$.)

Since any valid assemblage has a quantum realisation, it is trivial to write down a semi-definite program to find the quantum maximum $Q$ of $F$:
\begin{equation}
\begin{aligned}
& \text{maximize} & & \sum_{a,x} \tr(F_{a|x}\sigma_{a|x}) \\
& \text{subject to} & & \sigma_{a|x} \geq 0, \\
& & & \sum_a \sigma_{a|1} = \sum_a \sigma_{a|x} \ \forall x \in \{2, \dotsc, m_A\}, \\
& & & \sum_a \tr(\sigma_{a|1}) = 1.
\end{aligned}
\label{qmax}
\end{equation}

Now consider the cases when the assemblage has a LHS model. Notice that by shifting randomness into $p(\lambda)$, we can always make Alice's part of the model deterministic, i.e. let $\lambda : \{1, \dotsc, m_A\} \to \{1, \dotsc, n_A\}$ and $p(a|x,\lambda) = \delta_{a,\lambda(x)}$. We can furthermore combine $p(\lambda)$ and $\sigma_\lambda$ into subnormalized states $\tilde\sigma_\lambda = p(\lambda)\sigma_\lambda$. Hence an assemblage has an LHS model if and only if there exist ${n_A}^{m_A}$ positive $\tilde\sigma_\lambda$ with $\sum_\lambda \tr(\tilde\sigma_\lambda) = 1$ such that
\begin{equation}
\sigma_{a|x} = \sum_\lambda \delta_{a,\lambda(x)} \tilde\sigma_\lambda = \sum_{\substack{\lambda\\\lambda(x) = a}} \tilde\sigma_\lambda.
\end{equation}

With the above reformulation in hand, we can now write down a semi-definite program to find the maximum value $L$ of $F$ over all LHS models:
\begin{equation}
\begin{aligned}
&\text{maximize} & & \sum_\lambda \tr\left(\left(\sum_x F_{\lambda(x)|x}\right)\tilde\sigma_\lambda\right) \\
&\text{subject to} & & \tilde\sigma_\lambda \geq 0, \\
& & & \sum_\lambda \tr(\tilde\sigma_\lambda) = 1.
\end{aligned}
\label{smax}
\end{equation}

$F \leq L$ is called a (linear) steering inequality. If $L < Q$ then the inequality is non-trivial, i.e. can be violated in QM. More general (non-linear) steering inequalities have also been found, but I will not consider them here as they do not appear to be amenable to the techniques below. This restriction is not too onerous since every assemblage without an LHS model violates some linear steering inequality \cite{expcrit}.

\section{Bounding negativity}
If one observes an assemblage $\sigma_{a|x}$ that lacks a LHS model then one can conclude that it must have arisen from Alice measuring her half of some entangled state $\rho_{AB}$. We would now like to make that statement quantitative, i.e. find a lower bound on the amount of entanglement in $\rho_{AB}$. A lower bound is the best we can hope for, since Alice might ``wasted'' entanglement by choosing sub-optimal measurements. If we quantify entanglement by the negativity $N$ then we are trying to
\begin{equation}
\begin{aligned}
&\text{minimize} & & N(\rho_{AB}) \\
&\text{subject to} & & \tr_A((E_{a|x}\otimes \idn{B}) \rho_{AB}) = \sigma_{a|x}, \\
& & & \rho_{AB}, E_{a|x} \geq 0,\\
& & & \sum_a E_{a|x} = \idn{A} \ \forall x.
\end{aligned}
\label{naiveproblem}
\end{equation}

(We do not need to require $\rho_{AB}$ has unit trace since this follows from the normalization of $\sigma_{a|x}$.) This would appear to be a difficult problem, firstly because we need to consider all possible dimensions $d_A$ for Alice's system, and secondly because $(E_{a|x}\otimes \idn{B})\rho_{AB}$ contains the products of two unknowns $E_{a|x}$ and $\rho_{AB}$.

Adapting the techniques of Moroder \emph{et. al.} \cite{deq} (which are based on the ``NPA hierarchy'' \cite{hierarchy}), we can relax \eqref{naiveproblem} in a way that removes both difficulties. First notice that without loss of generality we can take the $E_{a|x}$ to be projective measurements, possibly by increasing $d_A$. Adopt the shorthand $A_0 = \idn{A}$, $A_1 = E_{1|1}$, $A_2 = E_{2|1}$, up to $A_{(n_A-1)m_A} = E_{n_A - 1 | m_A}$, i.e. $\{A_i\}$ consists of the identity plus all except the last $E_{a|x}$ for each setting $x$. Define a completely positive map on Alice's system $\chi(\rho_{AB}) = \sum_n (K_n\otimes I_B) \rho_{AB} (K_n^\dagger \otimes I_B)$ by $K_n = \sum_i \ket{i}\bra{n}A_i$. (The key difference from \cite{deq} is that here an anologous map is \emph{not} applied by Bob.) Then
\begin{equation}
 \chi(\rho_{AB}) = \sum_{ij} \ket{i}\bra{j} \tr_A((A_j^\dagger A_i \otimes \idn{B})\rho_{AB}).
\end{equation}
In fact there is an infinite hierarchy of relaxations, with the above being used at level $l=1$. In general, $\chi$ maps Alice's system to $l$ qu$d$its, with $d = (n_a-1)m_A + 1$, using $K_n = \sum_{i_1, \dotsc, i_l} \ket{i_1, \dotsc, i_l}\bra{n}A_{i_1} A_{i_2}\dotsm A_{i_l}$.

The basic idea is to optimize over possible $\chi(\rho_{AB})$ instead of $\rho_{AB}$ itself. Hence we need to translate each condition in \eqref{naiveproblem}. The first condition can be enforced using $\ket{i, 0,\cdots,0}\bra{0,\cdots,0}$ blocks of $\chi(\rho_{AB})$, which should be equal to $\tr_A( (A_i \otimes \idn{B})\rho_{AB})$. Since $\chi$ is completely positive we can relax $\rho_{AB} \geq 0$ to $\chi(\rho_{AB}) \geq 0$. The positivity of the measurement outcome is enforced by taking them to be projectors, and the final requirement of summing to identity has become implicit by not including the last outcome of each measurement in the $A_i$.

The form the map $\chi$ also puts several (linear) restrictions on $\chi(\rho_{AB})$, the satisfaction of which I will call ``$\chi$-validity''. Firstly since $A_0 = \idn{A}$ any blocks whose indices are the same when ignoring zeros must have identical contents. There are further constraints from the fact that the $A_i$ are hermitian, squares to themselves, and are orthogonal to other $A_j$ with the same setting. For example, if $n_A = 3$, $m_A=2$ and $l=1$ then we have the block-matrix form
\begin{equation}
\chi(\rho_{AB}) = \begin{pmatrix}
\sigma_\text{r} & \sigma_{1|1} & \sigma_{2|1} & \sigma_{1|2} & \sigma_{2|2} \\
\sigma_{1|1} & \sigma_{1|1} & 0 & X_1 & X_2 \\
\sigma_{2|1} & 0 & \sigma_{2|1} & X_3 & X_4 \\
\sigma_{1|2} & X_1^\dagger & X_3^\dagger & \sigma_{1|2} & 0 \\
\sigma_{2|2} & X_2^\dagger & X_4^\dagger & 0 & \sigma_{2|2}
\end{pmatrix},
\label{blockform}
\end{equation}
where $\sigma_r$ is Bob's reduced state $\sum_a \sigma_{a|x}$ and $X_i$ are arbitrary matrices (for example $X_1 = \tr_A( (E_{1|2}E_{1|1}) \rho_{AB})$ which is not an observable quantity). The reader may find it helpful to compare \eqref{blockform} with equation 7 of \cite{deq}.

The final step is to translate the objective function $N(\rho_{AB})$. Similarly to \cite{deq}, write $N(\rho_{AB}) = \min\{\tr(\rho_-) | \rho_{AB} = \rho_+ - \rho_-, \rho_\pm^{T_B} \geq 0\}$ and relax this to $\min\{t(\chi(\rho_-)) | \chi(\rho_{AB}) = \chi(\rho_+) - \chi(\rho_-), \chi(\rho_\pm)^{T_B} \geq 0\}$. $t(\chi(\rho))$ indicates the trace of the $\ket{0,\dotsc,0}\bra{0,\dotsc,0}$ block of $\chi(\rho_{AB})$, such that $t(\chi(\rho)) = \tr(\tr_A(\rho)) = \tr(\rho)$. Also, $\chi(\rho)^{T_B} = \chi(\rho^{T_B})$ since $\chi$ is local to Alice's system.

So the final form is
\begin{equation}
\begin{aligned}
&\text{minimize} & & t(\chi_-) \\
&\text{subject to} & & \chi_+ - \chi_- \text{ matches assemblage}, & \\
& & & \chi_+ - \chi_- \geq 0,\\
& & & \chi_\pm^{T_B} \geq 0,\\
& & & \chi_\pm \text{ are $\chi$-valid},
\end{aligned}
\label{relaxedproblem}
\end{equation}

whose solution, as argued above, lower bounds the solution of \eqref{naiveproblem}. If one is not interested in specific assemblage but rather given value $v$ of a steering functional $F$, then one should
\begin{equation}
\begin{aligned}
&\text{minimize} & & t(\chi_-) \\
&\text{subject to} & & f(\chi_+ - \chi_-) = v,\\
& & & t(\chi_+ - \chi_-) = 1,\\
& & & \chi_+ - \chi_- \geq 0,\\
& & & \chi_\pm^{T_B} \geq 0,\\
& & & \chi_\pm \text{ are $\chi$-valid},
\end{aligned}
\label{givenviolproblem}
\end{equation}
where $f(\cdot)$ is defined as the evaluation of $F$ using the appropriate blocks of $\chi$, i.e. $f(\chi(\rho)) = F(\rho)$. Finally, if one wants to upper bound the value of $F$ on states with no negativity (called positive partial transpose, or PPT, states), then one should
\begin{equation}
\begin{aligned}
&\text{maximize} & & f(\chi) \\
&\text{subject to} & & t(\chi) = 1,\\
& & & \chi \geq 0,\\
& & & \chi^{T_B} \geq 0,\\
& & & \chi \text{ is $\chi$-valid}.
\end{aligned}
\label{PPTproblem}
\end{equation}

\section{Results: stronger Peres conjecture?}
\begin{figure}
\includegraphics[width=\columnwidth]{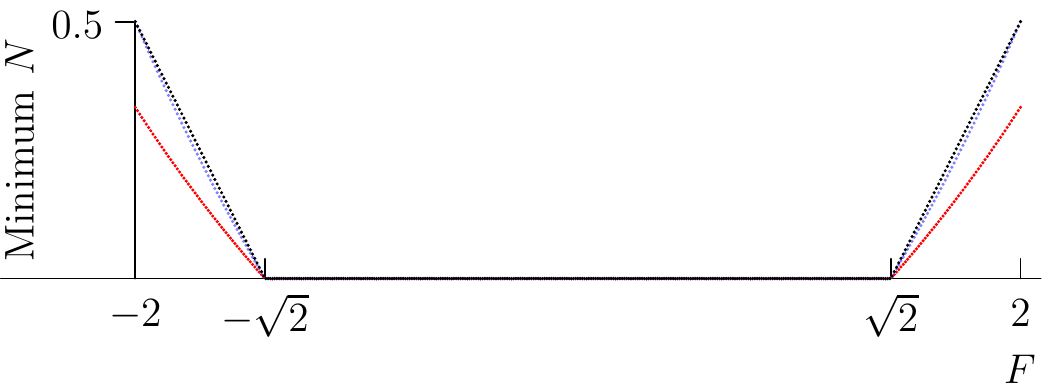}
\caption{The results of \eqref{givenviolproblem} applied to a simple steering inequality. The lowest (red) curve is $l=1$, the next (blue) is $l=2$, and the highest (black) is $l=3$.}
\label{res1}
\end{figure}

I implemented \eqref{qmax}, \eqref{smax}, \eqref{relaxedproblem}, \eqref{givenviolproblem} and \eqref{PPTproblem} in MATLAB using the YALMIP \cite{yalmip} modelling system. The scripts are available at \cite{website}. One of the simplest steering inequalities is equation 63 in \cite{expcrit}, which applies in the case $n_A = m_A = d_B = 2$ and in the present notation is proportional to $F_{1|1} = X$, $F_{2|1} = -X$, $F_{1|2} = Y$ and $F_{2|2} = -Y$ where $X$ and $Y$ are the Pauli matrices. LHS models satisfy $F \leq \sqrt{2}$ whilst the quantum maximum is $F = 2$. The results of \eqref{givenviolproblem} are shown in Fig~\ref{res1}.

Focusing on the $\sqrt2 \leq F \leq 2$ we see that at $l=3$ we have convergence to the bound $N \geq \frac{F - \sqrt{2}}{4 - 2\sqrt{2}}$. This bound is tight because $F = \sqrt{2}$ can be achieved with a separable state ($N=0$), whilst $F = 2$ can be achieved with a maximally entangled two-qubit state ($N=\frac12$). The points between can therefore be achieved by convex mixtures of the two, by the reasoning spelled out in the appendix of \cite{deq}.

\begin{figure}
\includegraphics[width=\columnwidth]{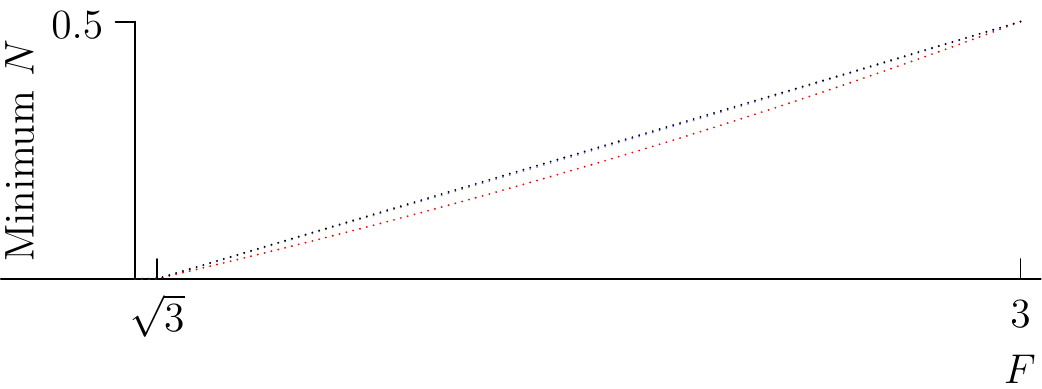}
\caption{The results of \eqref{givenviolproblem} applied to a another steering inequality. The lowest (red) curve is $l=1$, the next (blue) is $l=2$, and the highest (black) is $l=3$.}
\label{res2}
\end{figure}

A slightly more involved steering inequality, with $m_A = 3$, is equation 66 of \cite{expcrit}, which is obtained by adding $F_{1|3} = Z$ and $F_{2|3} = -Z$ to the previous case. Now $F \leq \sqrt{3}$ for LHS models whilst the quantum maximum is $F = 3$. The results of \eqref{givenviolproblem} for this inequality are shown in Fig~\ref{res2}. Notice that the Werner state $\rho_{0.6} = 0.6\ket{\psi_-}\bra{\psi_-} + 0.1 I$ (where $\ket{\psi_-} = (\ket{0}\ket{1} - \ket{1}\ket{0})/\sqrt2$) gives $F=1.8 > \sqrt{3}$. Hence the presence of negativity in that state can be certified, even though $\rho_{0.6}$ has a LHV model \cite{wernerLHV} and therefore no entanglement could be certified if neither party were trusted.

A common feature of both examples is that any $F$ outside the range of LHS models signifies the presence of negativity. This is somewhat surprising, since there are PPT (i.e. zero negativity) states that are nonetheless entangled \cite{PPTentangled}. It is \emph{prima facie} possible for such states to violate a steering inequality.

In the Bell scenario, Peres has conjectured \cite{peresconjecture} that the probabilities from measuring PPT states always have an LHV model, a conjecture supported by the results of \cite{deq}. Although a multi-partite version of this conjecture has been disproved \cite{multipartiteperes}, the bipartite case remains open. Based on the above observation, one might tentatively conjecture that PPT states cannot violate steering inequalities, i.e. the assemblages obtained by measuring them always have LHS models. Since an LHS model implies and LHV model, but not vice versa, this statement is strictly stronger than the original Peres conjecture. Hence if the original Peres conjecture is false, this strengthened conjecture may be a good starting point to seek counterexamples.

The methods provided in this paper can be used to search for counterexamples to this stronger conjecture. In that direction, I have used \eqref{PPTproblem} to upper bound the PPT violations of various steering inequalities. In all but one of the cases I have tried, an upper bound agreeing (within numerical precision) to the LHS bound is always found, supporting the strengthened conjecture. See Table~\ref{PPTres} for details. The exception was Eq. 1 of \cite{expEPR} with $n=10$. At the first level PPT bound is approximately 0.0537 above the LHS bound. At the second level the difference is approximately 0.0012. Unfortunately the third level is not tractable on my hardware, so the results for this inequality are inconclusive.

\begin{table}
\begin{tabular}{|l|l|l|l|l|}
  \hline
Inequality & $d_B$ & $m_A$ & $n_A$ & $l$ \\
  \hline
Eq. 63 of \cite{expcrit} & 2 & 2 & 2 & 1\\
Eq. 66 of \cite{expcrit} & 2 & 3 & 2 & 1\\
Eq. 67 of \cite{expcrit}, $j=1$ & 3 & 3 & 3 & 1\\
Eq. 67 of \cite{expcrit}, $j=3/2$ & 4 & 3 & 4 & 1\\
Eq. 67 of \cite{expcrit}, $j=2$ & 5 & 3 & 5 & 1\\
Eq. 67 of \cite{expcrit}, $j=5/2$ & 6 & 3 & 6 & 1\\
Eq. 67 of \cite{expcrit}, $j=3$ & 7 & 3 & 7 & 1\\
Eq. 67 of \cite{expcrit}, $j=7/2$ & 8 & 3 & 8 & 1\\
Eq. 67 of \cite{expcrit}, $j=4$ & 9 & 3 & 9 & 1\\
Eq. 14 of \cite{simplest} & 2 & 2 & 2 & 1\\
Eq. 1 of \cite{expEPR}, $n=4$ & 2 & 4 & 2 & 1\\
Eq. 1 of \cite{expEPR}, $n=6$ & 2 & 6 & 2 & 2\\
Eq. 1 of \cite{expEPR}, $n=10$ & 2 & 10 & 2 & see text\\
Eq. 7 of \cite{wierdmeas}, $n=4$ & 2 & 4 & 2 & 2\\
Eq. 7 of \cite{wierdmeas}, $n=5$ & 2 & 5 & 2 & 2\\
\hline
\end{tabular}
\caption{List of steering inequalities for which I have compared the ranges obtained by LHS models to the ranges obtained by PPT states. $d_B$ is the dimension of Bob's system, $m_A$ and $n_A$ are the number of settings and outcomes for Alice. The two ranges agree within numerical precision at level $l$ of the hierarchy of bounds on the PPT range.}
\label{PPTres}
\end{table}

All the steering inequalities in Table~\ref{PPTres} are fairly ``natural''/``symmetric'', and this might be a problem when searching for a counter-example to the strengthened Peres conjecture. Therefore I have also tried a different strategy of just randomly generating operators $F_{a|x}$, using \eqref{smax} to bound their values on LHS models and then comparing that with the bounds from \eqref{PPTproblem}. The limiting factor on increasing the parameters $d_A$, $n_A$, $n_B$ appears to be \eqref{smax}. In Table ~\ref{randres} I list the cases in which I was able to generate 4000 random sets of operators and check for counterexamples. None were found.

\begin{table}
\begin{tabular}{lll}
\begin{tabular}{|l|l|l|l|}
  \hline
$d_B$ & $m_A$ & $n_A$ & $l$ \\
  \hline
2 & 2 & 2 & 1\\
2 & 2 & 3 & 1\\
2 & 2 & 4 & 1\\
2 & 2 & 5 & 1\\
2 & 2 & 6 & 1\\
2 & 3 & 2 & 2\\
2 & 3 & 3 & 1\\
2 & 3 & 4 & 1\\
2 & 4 & 2 & 1\\
2 & 4 & 3 & 2\\
3 & 2 & 2 & 1\\
3 & 2 & 3 & 1\\
\hline
\end{tabular}
&\qquad \qquad  &
\begin{tabular}{|l|l|l|l|}
  \hline
$d_B$ & $m_A$ & $n_A$ & $l$ \\
  \hline
3 & 2 & 4 & 1\\
3 & 3 & 2 & 1\\
3 & 3 & 3 & 1\\
3 & 3 & 4 & 1\\
3 & 4 & 2 & 2\\
3 & 4 & 3 & 1\\
4 & 2 & 2 & 1\\
4 & 2 & 3 & 1\\
4 & 2 & 4 & 1\\
4 & 3 & 2 & 1\\
4 & 3 & 3 & 1\\
4 & 4 & 2 & 1\\
\hline
\end{tabular}
\end{tabular}
\caption{List of parameter regimes for which I have generated 4000 random steering inequalities and checked for counterexamples to the stronger Peres conjecture. The final column shows the level of the hierarchy at which agreement between \eqref{smax} and \eqref{PPTproblem} was achieved to within numerical precision for the ``hardest'' inequality in that regime.}
\label{randres}
\end{table}

\section{Conclusions}

The EPR-steering scenario is an interesting middle ground in which to study entanglement. The entanglement of some states, invisible in the fully device independent scenerio due to the existence of an LHV model, can be quantified using the techniques described above. On the other hand, there are certainly entangled states that have LHS models \cite{steer}, so some entanglement can only be quantified when both parties are trusted. It appears to be possible that all PPT entangled states are in the latter category. This is a stronger version of the Peres conjecture, and is the main open question posed here.

A more technical question I have not addressed is whether the methods here can be proven to always converge to a tight bound, as was shown for \cite{deq}.

Finally, a more conceptual open question is whether the EPR-steering scenario allows for the quantification of anything other than negativity. It would be particularly interesting if that were possible for a quantity that is completely unavailable in the fully device independent scenario.

\begin{acknowledgements}
Thanks to Peter Lewis and David Jennings for discussions, and to Eric Calvalcanti for useful discussions. I acknowledge financial support for the EPSRC.
\end{acknowledgements}

\bibliography{steer}

\end{document}